\documentclass[reqno,12pt]{article}
\usepackage{amsmath,amsfonts,amssymb,amsthm,amstext,amscd,eucal,xcolor}
\usepackage[all]{xy}
\usepackage{hyperref}
\usepackage{epsfig}
\usepackage[active]{srcltx}
\makeatletter \@addtoreset{equation}{section}

\makeatletter\renewcommand\section{\@startsection {section}{1}{\z@}%
                                   {-3.5ex \@plus -1ex \@minus -.2ex}
                                   {2.3ex \@plus.2ex}%
                                   {\normalfont\large\bfseries}}
\renewcommand\subsection{\@startsection{subsection}{2}{\z@}%
                                     {-3.25ex\@plus -1ex \@minus -.2ex}%
                                     {1.5ex \@plus .2ex}%
                                     {\normalfont\bfseries}}

\newcommand{\be}{\begin{equation}}
\newcommand{\ee}{\end{equation}}
\newcommand{\bea}{\begin{eqnarray}}
\newcommand{\eea}{\end{eqnarray}}
\newcommand{\bse}{\begin{subequations}}
\newcommand{\ese}{\end{subequations}}
\newcommand{\beqa}{\begin{eqnarray}}
\newcommand{\eeqa}{\end{eqnarray}}
\newcommand{\beqar}{\begin{eqnarray*}}
\newcommand{\eeqar}{\end{eqnarray*}}
\newcommand{\bi}{\begin{itemize}}
\newcommand{\ei}{\end{itemize}}
\newcommand{\bn}{\begin{enumerate}}
\newcommand{\en}{\end{enumerate}}

\newcommand{\ba}{\begin{array}}
\newcommand{\ea}{\end{array}}
\newcommand{\bc}{\begin{center}}
\newcommand{\ec}{\end{center}}
\newcommand{\nnr}{\nonumber \\}

\def\lemd{EMD-$\Lambda$ }

\newcommand{\ads}[1]{AdS$_{#1}$}

\definecolor{darkgreen}{rgb}{0,0.3,0}
\definecolor{darkblue}{rgb}{0,0,0.3}
\definecolor{darkred}{rgb}{0.7,0,0}

\begin{document}

\begin{titlepage}

\begin{flushright}\vspace{-3cm}
{\small
IPM/P-2014/nnn \\
\today }\end{flushright}
\vspace{0.5cm}

\begin{center}
\centerline{{\Large{\bf{On  Classification of Geometries with SO(2,2) Symmetry}}}} \vspace{6mm}

\large{\bf{S. Sadeghian\footnote{e-mail: s\_sadeghian@alzahra.ac.ir}$^{;\ a,b}$, M.M. Sheikh-Jabbari\footnote{e-mail:
jabbari@theory.ipm.ac.ir}$^{;\ a}$
 and  H. Yavartanoo\footnote{e-mail:
yavar@itp.ac.cn }$^{;\ c}$  }}
\\

\vspace{5mm}
\normalsize
\bigskip\medskip
{$^a$ \it School of Physics, Institute for Research in Fundamental
Sciences (IPM),\\ P.O.Box 19395-5531, Tehran, Iran}\\
\smallskip
{$^b$ \it Department of Physics, Alzahra University P. O. Box 19938, Tehran 91167, Iran}\\
\smallskip
{$^c$ \it  State Key Laboratory of Theoretical Physics, Institute of Theoretical Physics,
Chinese Academy of Sciences, Beijing 100190, China.
 }\\

\end{center}

\begin{abstract}
\noindent
Motivated by the Extremal Vanishing Horizon (EVH) black holes, their near horizon geometry and the EVH/CFT proposal \cite{EVH/CFT}, we construct and classify solutions with (local) $SO(2,2)$ symmetry to four and five dimensional Einstein-Maxwell-Dilaton (EMD) theory with positive, zero or negative cosmological constant $\Lambda$, the EMD-$\Lambda$ theory, and also $U(1)^4$ gauged supergravity in four dimensions and $U(1)^3$ gauged supergravity in five dimensions.
In four dimensions the geometries are warped product of \ads{3} with an interval or a circle. In five dimensions the geometries are of the form of warped product of \ads{3} and a 2d surface $\Sigma_2$. For the Einsten-Maxwell-$\Lambda$ theory we prove that $\Sigma_2$ should have a $U(1)$ isometry, a rigidity theorem in this class of solutions. We also construct all $d$ dimensional Einstein vacuum solutions with $SO(2,2)\times U(1)^{d-4}$ isometry.

\end{abstract}


\end{titlepage}
\setcounter{footnote}{0}
\renewcommand{\baselinestretch}{1.05}  

\addtocontents{toc}{\protect\setcounter{tocdepth}{2}}
\tableofcontents


\section{Introduction and Motivation}\label{sec-intro}

Construction and classification of solutions to Einstein gravity with various kind of matter fields in diverse dimensions has been an interesting question under intense study since the conception of general relativity e.g. see \cite{Hawking-Ellis, Emparan-Reall-review} and references therein. Black hole, black ring, black brane or string solutions, geometries which are stationary and have Killing and/or event horizon, have been of particular interest. These ``black geometries'' are generically specified by their conserved Noether-Wald  \cite{Wald} and possibly other \cite{Other-charges} charges  (mass, angular momenta and electric or magnetic or higher-form charges, dipole charges and ...). These black geometries may exist in asymptotic flat, de Sitter or anti-de Sitter space. To specify the solutions besides the asymptotic behavior and charges of the solutions, among other things, one needs to also specify  the topology of the horizon manifold, which is a codimension 2d surface, e.g. see \cite{Emparan-Reall-review}.

In this work we will focus on the Einstein-Maxwell-Dilaton theory, in presence of a generic cosmological constant term $\Lambda$, the EMD-$\Lambda$ theory, in four and five dimensions and explore a specific class of its solution which are not asymptotically (locally) $R^d$, dS$_d$ or AdS$_d$ ($d=4,5$). The solutions of our interest have a (globally defined) time-like Killing vector field but are not necessarily black holes. We will construct and classify  solutions with local $SO(2,2)$ isometry to four and five dimensional EMD-$\Lambda$ theory.

The geometries of our interest are hence those with a (local) \ads{3} factor. Our interest in these geometries is primarily motivated by the fact that there is a special class of extremal black holes, Extremal Vanishing Horizon (EVH) black holes, where in the near horizon limit lead to such geometries, see \cite{EVH/CFT,EVH-2,EVH-3,AdS4-EVH,Japanese-EVH,EVH-Ring} for previous analysis of EVH black holes. This should be contrasted with the near horizon limit of usual extremal black holes, where one generically finds an \ads{2} factor, rather than an \ads{3} \cite{KL-papers,{KL-review}}. The appearance of local \ads{3} factor in the EVH case can be attributed to the fact that for EVH black holes  the co-dimension two horizon surface of the black hole horizon has the peculiar property that its area vanishes due to the presence of a vanishing one-cycle on the horizon \cite{First-law}. In the near horizon limit this vanishing one-cycle joins with the \ads{2} directions one would normally expect in the near horizon of extremal black holes, to make an \ads{3} \cite{First-law}.
One may then argue, based on presence of this \ads{3} factor in the near horizon limit, for a dual 2d CFT description for near-EVH (low energy excitations of EVH) black holes, the EVH/CFT correspondence \cite{EVH/CFT}.

Due to the fact that generic gravity theories do have EVH black hole solutions (based on the analysis in the papers mentioned in the last paragraph) \footnote{
EVH black holes can exist in any dimension, in fact massless BTZ is the simplest EVH black hole \cite{BTZ-EVH}. Moreover, geometries with \ads{3} factors, may also arise in Ricci-flat reductions of (super)gravity theories, as recently discuss in \cite{Flat-reduction}.}
one is naturally led to the question of construction and classification of geometries with local $SO(2,2)$ isometry. This is what we will tackle in this work.

In section \ref{sec-2}  we construct and classify all geometries with $SO(2,2)$ isometry in four dimensional  \lemd theory. We first construct vacuum solutions, then consider cases with cosmological constant $\Lambda$. The $SO(2,2)$ invariance forbids four dimensional Maxwell fields ($U(1)$ gauge fields) to have a non-zero field strength and hence the only case to consider here is addition of scalar (or dilaton) fields and to construct the class of solutions for this case.

In section \ref{sec-3} we consider the five dimensional case where we prove a ``rigidity theorem'': all five dimensional solutions to Einstein-Maxwell-$\Lambda$ theory with local $SO(2,2)$ isometry also exhibit an extra $U(1)$ isometry. The $SO(2,2)$ corresponds to isometries along an \ads{3} part while the $U(1)$ corresponds to rotations on the other two dimensional part. We construct all  $SO(2,2)$ invariant solutions of this theory. Addition of the dilaton and/or scalar to this theory, we construct $SO(2,2)$ invariant solutions, assuming this extra $U(1)$ isometry. We also construct a large class of $SO(2,2)\times U(1)$ invariant solutions to five dimensional $U(1)^3$ gauged supergravity.

In section \ref{sec-4} we construct all $SO(2,2)\times U(1)^{d-4}$ invariant vacuum solutions to Einstein gravity in generic dimension $d$ and briefly discuss other possible solutions with less number of $U(1)$ isometries.

Section \ref{sec-5} is devoted to concluding remarks. In an appendix we discuss $d\to d-3$ reduction and briefly the $d\to 3$ reduction which may be respectively facilitate construction of the solutions and the dual 2d CFT description associated with the solutions.


\section{4d solutions}\label{sec-2}
In this section we study $SO(2,2)$ invariant solutions of Einstein theory of gravity coupled with scalar and Maxwell gauge fields, described by the action
\be\label{EMS-action}
S=\frac{1}{16\pi G_N}\int d^4x \sqrt{-g} \left(R- \sum_I\ e^{-\alpha_I^i\Phi_i} F_I^2+ G^{ij}(\Phi)g^{\mu\nu}\partial_{\mu}\Phi_i\partial_{\nu}\Phi_j -V(\Phi)\right)
\ee
where $G_{ij}=G_{ij}(\Phi)$ is the metric on the space of scalar fields and $F_I=dA_I$ are the $U(1)$ gauge field strengths.
The most general form of a four dimensional metric with $SO(2,2)$ isometry can be written as
\bea \label{4Dmetric}
ds^2=e^{f(\theta)} (ds_3^2+\beta^2d\theta^2), \quad
\eea
where $ds_3^2$ denotes the locally AdS$_{3}$ metric of unit radius:
\be\label{ds3}
ds_3^2=-r^2dt^2+\frac{dr^2}{r^2}+r^2d\varphi^2,
\ee
and $\beta$ is a constant which determines periodicity or range of the $\theta$ coordinate, for solutions in which $\theta$ is periodic or bounded.  The $SO(2,2)$ invariance manifests itself in the Ricci curvature of the metric through the fact that $R_{\theta\theta}$,  $R_{\theta a}$ and $R_{ab},\ a,b=r,t,\varphi$ components of Ricci are respectively in scalar, 3 and 6 irreps of $SO(2,2)$.

Since $SO(2,2)$ does not have invariant two-forms, $SO(2,2)$ invariance forbids $U(1)$ gauge fields to have non-vanishing field strength. Moreover, the scalar fields $\Phi_i$ cannot have $r,t,\varphi$ dependence, because the energy momentum tensor should fall into the same $SO(2,2)$ representation as the Ricci tensor described above.

For specific solutions, let us focus on the case with one scalar $\Phi$, cases with more scalars generically appear as straightforward generalisations.
If we take scalar coupling $G(\Phi)=-2$, equations of motion for scalar fields and the metric are given by
\be
R_{\mu\nu}=\frac{1}{2}g_{\mu\nu} V+2 \partial_{\mu}\Phi  \partial_{\nu}\Phi, \quad  \nabla^2\Phi=\frac{1}{4}\frac{dV}{d\Phi}.
\ee
Noting the metric ansatz \eqref{4Dmetric} and that the scalar field can be only a function of $\theta$,
equations of motion reduce to
\bea\label{4d-scalar-EOM}
&&  4\beta^2+f^{'2}-2f^{''}-4\Phi^{'2}=0, \\
&& 12\beta^2+3f^{'2}-4\Phi^{'2}+2\beta^2 V e^f=0, \\  \label{4d-scalar-EOM2}
&&  \Phi^{''}+\Phi' f' -\frac{1}{4}\beta^2 e^f \frac{dV}{d\Phi}=0
\eea
One can readily check that, as expected, the above three equations for two unknown functions $f, \Phi$ are compatible with each other. The above second order differential equations may be solved for a given scalar potential V.

As the starter, we show that there are no four dimensional $SO(2,2)$ invariant vacuum solution with $R_{\mu\nu}=0$. To this end, we turn off scalars and their potential and the equations of motion reduce to
\be
f^{''}=0, \qquad f^{'2}+4\beta^2=0
\ee
which does not have any real solution. We conclude such geometry does not exist in the pure Einstein gravity. As we will see in section 2.2, addition of cosmological constant may change this result. It is interesting to analyse the effect of higher derivative correction to this conclusion.

\subsection{Dilaton case}

When the potential $V(\Phi)$ vanishes, the action \eqref{EMS-action} has a shift symmetry for scalars  and we are dealing with the EMD case. In this case the equations reduce to
\be
\Phi'=\lambda e^{-f}\,,\qquad 12\beta^2+3f'^2=4\lambda^2e^{-2f}\,,
\ee
where $\lambda$ is an integration constant. The most general solutions of above equations are given by:
\begin{itemize}
\item[I.] \ads{3}$\times $S$^1$ solution
\be\label{ads3s1}
e^f=\frac{\lambda}{\sqrt3 \beta}=const. \,,\qquad \Phi'={\sqrt3 \beta}\,,
\ee
with \ads{3} radius $R^2_3=\frac{\lambda}{\sqrt3 \beta}$ and S$^1$ radius $\beta R_3$, if we  choose $\theta\in [0,2\pi]$. This \ads{3}$\times $S$^1$ solution may be thought as a linear dilaton (four dimensional) non-critical string theory background, similar to the solutions discussed in \cite{Klebanov-Maldacena}.

\item[II.] Near horizon of 4d EVH black holes \cite{EVH/CFT}

\be\label{NH4dEVH}
e^f=\frac{\lambda}{\sqrt3\beta} \sin 2\beta\theta,\quad  \Phi = \frac{\sqrt{3}}{2}\ln (\tan \beta\theta )\,.
\ee
One may conveniently choose $\beta=1$. Note that it was proved in \cite{EVH/CFT} that any EVH black hole solution to 4d EMD theory has the same near horizon limit, specified by the above functions, with $\theta\in [0,\pi]$ range. In the $\theta=\pi/2$ the dilaton blows up and the appropriate description of the geometry is given through a 5d uplift of the geometry. In the 5d case the solution corresponds to the near horizon limit of EVH KK black hole \cite{EVH/CFT}. We will return to this 5d solution in the next section.\footnote{Since the unknown functions in the metric have only $\theta$ dependence, one may use a coordinate system in which the dilaton $\Phi$  is equal to $\theta$ itself, i.e. $\Phi'=\sqrt{3}\beta$. In this coordinate system metric takes the form
$ds^2=\frac{R^2}{\cosh 2\beta\theta} \left( ds_3^2 + \frac{\beta^2  d\theta^2 }{\cosh^2 2\beta\theta} \right).
$ }

\end{itemize}

\subsection{Solution at the extremum of $V(\Phi)$}
Although it is expected that  \eqref{4d-scalar-EOM} should have solutions for a generic $V(\Phi)$, finding explicit solutions is a formidable task. One specific solution which is usually of interest in such cases is when $V(\Phi)$ has an extremum (minimum). Let us denote this minimum by $\Phi_0$:
\be
\frac{dV}{d\Phi}\big|_{\Phi_0}=0\,,\qquad V(\Phi_0)=V_0\,.
\ee
The equations then immediately imply that $\Phi_0$ is a constant and hence the problem reduces to having Einstein gravity with a cosmological constant $\frac12 V_0$ (which is obviously  the same as the case with a constant potential $V_0$). The equation for metric coefficient $f$ takes the form
\be
U'^2+\beta^2 U^2+\frac16 V_0{\beta^2}=0\,,\qquad U=e^{-\frac12 f}\,.
\ee
The above equation has solution only if $V_0$ is negative and its solution is
\be
e^f=-\frac{6}{V_0} \sec^2\beta\theta\,.
\ee
With this $e^f$ and $\theta\in [0,\pi/\beta]$, \eqref{4Dmetric} is nothing but  \ads{4} of radius $\sqrt{\frac{-6}{V_0}}$ in an \ads{3} slicing.

\paragraph{The case of $U(1)^4$ gauged supergravity.} We now briefly discuss $SO(2,2)$ solutions of 4d $U(1)^4$ gauged supergravity. This theory has four Maxwell fields and three scalars with a given potential e.g. see \cite{ten-people}, and it falls into the class of theories described by the action \eqref{EMS-action}. As discussed, $SO(2,2)$ invariance of the solution does not allow for turning on gauge fields and hence we are left with a theory with three scalars $\Phi_i$ minimally coupled to Einstein gravity, with potential $-\frac{2}{L^2} \sum_i \cosh\Phi_i$. In this case the three scalars are coupled to each other only through back-reacting on the metric. There is an obvious solution with $\Phi_i=0$ with \ads{4} metric of radius $L$.

As the next example let us consider the case where the three scalars are related to each other so that we remain with an effective single scalar $\Phi$ with potential
\be
V=-(V_1 e^{\frac{2}{\sqrt{3}} \Phi}+V_2e^{-\frac{2}{\sqrt{3}} \Phi})\;.
\ee
It is easy to check that, equations of motion (\ref{4d-scalar-EOM})-(\ref{4d-scalar-EOM2}) admit following solution\footnote{This is a solution, not necessarily the most general solution.}
\be\label{2.16}
\Phi=\frac{\sqrt{6}\beta}{2} \theta,\qquad e^{-\frac{f}{2}}=\sqrt{\frac{V_1}{6}}\; e^{\frac{\beta\theta}{\sqrt{2}}} +\sqrt{\frac{V_2}{6}}\; e^{-\frac{\beta\theta}{\sqrt{2}}}
\ee

For the special case of $V_1=V_2=\frac{3}{2R^{2}}$, which is  obtained in the sector of $U(1)^4$ gauged SUGRA where $\Phi_i=\frac{2}{\sqrt{3}}\Phi$, and choosing $\beta=\sqrt2$,  this solution reduces to
\bea\label{2.17}
ds^2=\frac{R^2}{\cosh^2{\theta}}(ds^2_{3}+2 d\theta^2)=R^2\left(\cos^2{\alpha} ds_3^2+ 2d\alpha^2\right),\qquad \alpha \in [0,\pi/2]
\eea
where $ds^2_3$ is the (locally) \ads{3} metric of radius one.
\section{5d solutions}\label{sec-3}
As the four dimensional solutions indicate, the $SO(2,2)$ invariance implies that a three dimensional part of metric is fixed and that other component of metric (as well as gauge field and scalars) cannot depend on this three dimensional part. As such, as we go to higher dimensions there are much more freedom and more solutions. This is similar to black object solutions, where in 5d we do not generically have the uniqueness theorems we have in 4d, e.g. see \cite{ER-review, KL-new}. Moreover, unlike the 4d case $SO(2,2)$ invariance does not imply that gauge fields should be turned off. In this section we study five dimensional solutions in two derivative theories. Again we mainly focus on the Einstein-Maxwell-Dilaton theory (with possible addition of a potential for scalar field, as in the gauged supergravities).
 Let us start with the following action
\be
 S=\frac{1}{16\pi G_N}\int d^5x \sqrt{-g} \left(R+G_{mn}(\Phi)\partial_{\mu}\Phi^{m}\partial^{\nu}\Phi^{n} -f_{IJ}(\Phi) \; F^{I}_{\mu\nu}F^{J\mu\nu} -V(\Phi)\right).
\ee
As in the gauged SUGRAs the action may also involve  a Chern-Simons term. We will return to this later when we discuss $U(1)^3$ gauged SUGRA.

 Equations of motion for metric, scalar and gauge fields are given by
\begin{subequations}\label{5d-EOM}\begin{align}
R_{\mu\nu}+G_{mn} \partial_{\mu}\Phi^{m} \partial_{\nu}\Phi^{n} &=f_{IJ}(2F^{I}_{\mu\lambda} F_{\nu}^{J\lambda}-\frac{1}{3}g_{\mu\nu} F_{\alpha\beta}^{I} F^{J\alpha\beta})+\frac{1}{3}g_{\mu\nu} V, \\
 \frac{1}{\sqrt{-g}} \partial_{\mu}\left(\sqrt{-g} G_{mn} \partial^{\mu} \Phi^{n}\right) &=\frac{1}{2}\left(-\frac{\partial f_{IJ}}{\partial \Phi^{m}} F_{\mu\nu}^{I} F^{J\mu\nu}    + \frac{\partial G_{kl}}{\partial \Phi^{m}}\partial_{\mu}\Phi^{k}\partial^{\mu}\Phi^{l} -\frac{\partial V }{\partial \Phi^{m}} \right),\\
\partial_{\mu}\bigg[ \sqrt{-g} f_{IJ} F^{J\mu\nu}\bigg]&=0.
\end{align}\end{subequations}
The most general form of a five dimensional metric with $SO(2,2)$ isometry can be written as
\bea\label{genericmetric}
ds^2=e^{2R} ds_3^2 +e^{-R} g_{ij}dx^i dx^j,
\eea
where $R=R(\theta,\psi), g_{ij}=g_{ij}(\theta,\psi)$ and $i, j =\theta, \psi$ are coordinates of internal space.  We can still use two diffeomorphisms along $\theta,\psi$ and $g_{ij}$ will hence contain one unknown function. It appears that for different cases it is more convenient to choose this extra diffeo's accordingly. So, generically metric has two unknown functions.

$SO(2,2)$ invariance restricts the gauge field strength to be of the form
\be\label{gaugefield5dim}
F^{I}=F^{I}(\theta, \psi)\; d\theta \wedge d\psi\;,
\ee
and the scalars $\Phi_m=\Phi_m(\theta,\psi)$.

In this work we restrict to scalars with canonical kinetic term and set $G_{mn}(\Phi)=-2\delta_{mn}$. Moreover, except for section \ref{section-u1-3} where discuss $U(1)^3$ gauged SUGRA case, we consider cases with one scalar and one Maxwell field.

\subsection{Pure Einstein gravity}

Let us start with the simplest case of pure Einstein gravity. In this case, it is convenient to choose the 2d diffeos such that the 5d metric takes the form
\be\label{5d-vacuum}
ds^2=e^{2R} (ds^2_3+d\theta^2) +e^{2h}d\psi^2\,,
\ee
where  $R, h$ are two functions of $\theta,\psi$.

\paragraph{Theorem.} All $SO(2,2)$ invariant 5d vacuum Einstein solutions have necessarily an extra $U(1)$ isometry.
(This  theorem is somehow similar to the 4d rigidity theorem that any stationary vacuum asymptotic flat solution is also axisymmetric.)

\paragraph{Proof.} One may readily write down Einstein equations. Focusing on the $\theta\psi$ component of Einstein equations we learn that
\be
\partial_\psi R=0\,,\qquad \textrm{or}\qquad  \partial_\psi R=C'(\psi)e^{h}\,,
\ee
where $C'(\psi)$ is (derivative of) an arbitrary function of $\psi$ arising from integration over $\theta$ direction. If $\partial_\psi R=0$, then one immediately learns from other equations that $\partial_\psi h=0$ and hence $\partial_\psi$ provides a $U(1)$ Killing vector.

For the $\partial_\psi R=C'(\psi)e^{h}$ case,
one may absorb $C(\psi)$ in the part of diffeos which has not been used, redefining $\psi\to C(\psi)$, we end up with $\partial_\psi R=e^{h}$.
Next, we consider $\theta\theta$ and $\psi\psi$ components of the equations which lead to
\be\label{R-theta-psi-U(1)}
\frac{\partial_\psi R}{\partial_\theta R}=f'(\psi)\,,
\ee
where $f(\psi)$ is derivative of an arbitrary function of $\psi$. From the above one learns that $R=R(\theta+f(\psi))$. One may next redefine $\theta$ as $\theta+f(\psi)$. Therefore, functions $h$ and $R$ will be both only functions of this new coordinate $\theta$. That is, after this shift in $\theta$ coordinate, $\partial_\psi$ is a Killing direction. We note that this redefinition in $\theta$ coordinate introduces a $\theta\psi$ component in the metric. Nonetheless, noting that the metric components only depend on $\theta$, this off-diagonal component can be reabsorbed in a shift of $\psi$. So, we have proved that solutions to vacuum Einstein equations will be of the form \eqref{5d-vacuum} with $R=R(\theta), h=h(\theta)$ and $\psi$ is along a Killing direction.

In the above argument we only used two of the three independent Einstein equations. We can use the last one and explicitly solve for $R(\theta)$. The metric is then obtained to be
\be\label{Pure-5d-Einstein-soln}
ds^2=\ell^2\cos^2\theta ds^2_3 +\ell^2(\cos^2\theta d\theta^2+\tan^2\theta d\psi^2)\,,
\ee
where $ds^2_3$ is the metric for a locally \ads{3} geometry. We have chosen the integration constants such that $\theta\in[0,\pi/2]$ and $\psi\in[0,2\pi]$ and that the $\theta\psi$ part of metric does not have conic singularity. The two dimensional $\theta,\ \psi$ part of the metric is a compact finite volume.

It is important to note that this solution is not a regular solution and Kretschmann invariant  blows up at  $\theta=\pi/2$. Nonetheless, one may perform a reduction of the gravity theory over the $\theta\psi$ part and observe that one obtains, as expected, a 3d Einstein gravity with negative cosmological constant. If the Newton constant of the 5d theory is denoted by $G_5$ the 3d Newton constant is $G_3=G_5/(\pi \ell^2)$ and the \ads{3} radius is $\ell$ \cite{EVH-Ring}. One may then use \ads{3}/CFT$_2$ duality and propose that low energy effective dynamics of 5d gravity over the metric \eqref{Pure-5d-Einstein-soln} is described by a 2d CFT at central charge $c=3\ell/(2G_3)=3\pi \ell^3/(2G_5)$.

It is notable that the above metric is the one obtained in the near horizon of EVH Myers-Perry black holes \cite{EVH-Ring} (these are 5d Myers-Perry black holes \cite{Myers-Perry} with one spin set to zero) or, near horizon limit of balanced or unbalanced EVH black rings  \cite{EVH-Ring}. The latter two are the only known asymptotic flat vacuum solutions to 5d Einstein gravity.


\subsection{Einstein-$\Lambda$ theory}
It is straightforward to generalize above theorem and argument when the theory has a cosmological constant and  show that the geometry has an extra $U(1)$ isometry also in presence of $\Lambda$. Since the proof is essentially the same as the case without $\Lambda$, we do not repeat it here. Considering the $U(1)$, we adopt the metric ansatz
\be
\label{5dU1metric}
ds^2=e^{2R}(ds_3^2+d\theta^2)+e^{2h}d\psi^2\,,
\ee
with $R=R(\theta),\ h=h(\theta)$.

A combination of the Einstein equations then reduce to
\be\label{R-h-5d}
R''-R'h'=0 \quad \Longrightarrow  \quad  R'=\lambda e^h,
\ee
where $\lambda$ is an arbitrary integration constant. The rest of equations of motion reduce to the following equation
\be\label{R-Lambda-5d}
R''+R^{'2}+\frac{\Lambda}{3} e^{2R} +1=0\,.
\ee

For $R'=0$ case, \eqref{R-h-5d} is trivially satisfied, while \eqref{R-Lambda-5d} has solution only when $\Lambda<0$. For this special case, the equations of motion leads to the following equation for $h$
\be
h''+h'^2=1\,, \qquad R'=0\,,\qquad \Lambda<0\,.
\ee
Below we consider $R'\neq 0$ and $R'=0$ cases separately.

\paragraph{$\mathbf{R'\neq 0}$ case.} Equation \eqref{R-Lambda-5d} is the Jacobi elliptic equation for $e^R$. However, for explicit representation of metric,  it is more convenient to make a coordinate transformation on $\theta$ and write the metric into the following form
\be\label{5d-Lambda-metric}
ds^2=\ell^2 \cos^2\theta ds_3^2+\frac{a^2\cos^2\theta}{\Delta(\theta)} d\theta^2+\frac{a^2}{(1+\frac{\Lambda a^2}{6})^2}\Delta(\theta)\tan^2\theta d\psi^2\,,
\ee
where
\be
\Delta(\theta)=1+\frac{\Lambda a^2}{6}\cos^2\theta\,,\qquad \ell^2=\frac{a^2}{1-\frac{\Lambda a^2}{6}}\,,\qquad -1<\frac{\Lambda a^2}{6}<1\,.
\ee
In the above, $\theta\in [0,\pi/2]$ and to avoid conical singularity (at $\theta=0$), $\psi\in[0,2\pi]$.
This class of solutions exist for both negative and positive cosmological constant $\Lambda$. The metric \eqref{5d-Lambda-metric}, as expected, reduce to \eqref{Pure-5d-Einstein-soln} in the $\Lambda=0$.
We also comment that metric \eqref{5d-Lambda-metric} may also be obtained as the near horizon limit of EVH AdS-Myers-Perry. (This latter may be explicitly seen from eq.(4.18) of \cite{EVH-3} by setting $q=0$.)

One may reduce the Einstein-$\Lambda$ theory over the two dimensional $\theta\psi$ part to obtain a 3d Einstein gravity with cosmological constant $-2/\ell^2$ and the 3d Newton constant $G_3=G_5\frac{1+\Lambda a^2/6}{\pi a^2}$. This 3d theory can have a 2d CFT dual with Brown-Henneaux central charge $c=\frac{3\pi a^3}{2G_5}\left[(1+\Lambda a^2/6)\sqrt{1-\Lambda a^2/6}\right]^{-1}$.

AdS$_{5}$ geometry is the only other possible solution in this case which in the coordinate system (\ref{5dU1metric}) can be written as follows\footnote{This AdS$_5$ solution may formally be obtained  in the $a\to\infty$ limit ($\Lambda$ fixed) of (3.13), upon an extra Wick rotation on the $\theta$ coordinate.}  
\be \label{AdS5-AdS3-slice}
ds^2=\ell^2\left[\frac{1}{\cos^2\theta}(ds_3^2+d\theta^2+\sin^2\theta d\psi^2)\right]\,,\qquad \ell^2=-\frac{6}{\Lambda}\,,
\ee
where the $\theta\psi$ part is conformal to an $S^2$.
\paragraph{$\mathbf{R'=0}$ case.} In this case the solution turns out to an AdS$_3\times$H$^2$ geometry
\bea\label{special1}
ds^2=\ell^2 \left[ds_3^2+ \frac12 (d\theta^2+\sinh^2\theta d\psi^2)\right]\,,\qquad \ell^2=-\frac{3}{\Lambda}\,,
\eea
where  to avoid conic singularity we take $\psi\in [0,2\pi]$.
We also note that solutions (\ref{AdS5-AdS3-slice}) and (\ref{special1}) are not continuously connected to the vacuum one at $\Lambda=0$ discussed in the previous subsection, i.e. metric \eqref{Pure-5d-Einstein-soln}.

\subsection{5d Einstein-Maxwell case}
In this part we consider 5d Einstein-Maxwell-$\Lambda$ theory with the action
\be
 S=\frac{1}{16\pi G_N}\int d^5x \sqrt{-g} \left(R- \frac{3}{2}F_{\mu\nu}F^{\mu\nu} -2\Lambda\right)\,.
\ee
In five dimensions, $SO(2,2)$ invariance allows for having invariant two forms along the directions transverse to the \ads{3} part, i.e. a magnetic two form field strength.\footnote{We can of course also have $SO(2,2)$ invariant electric three forms, which is beyond the scope of the EMD theory we are studying in this work.} We start with extending the theorem for vacuum solutions to the Einstein-Maxwell-$\Lambda$ case.

\paragraph{Theorem.} All $SO(2,2)$ invariant solutions to 5d Einstein-Maxwell-$\Lambda$ theory have necessarily an extra $U(1)$ isometry.

\paragraph{Proof.} Let us start with metric and gauge field ansatzs  (\ref{5dU1metric}) and \eqref{gaugefield5dim}. Equations of motion for gauge fields can be solved to get
\be
\label{gaugefield5d}
F_{\theta\psi}=p \; e^{h-2R}\,.
\ee
With this gauge field strength one may readily examine the other equations of motion given in \eqref{5d-EOM}. Noting the form of gauge field strength, it is straightforward to verify that the steps of the proof discussed for the pure Einstein case goes through verbatim and we hence do not repeat it again here. We also note that similar discussion can be made for the case when there are more than one Maxwell fields; our ``$SO(2,2)$ rigidity theorem'' extends to cases with arbitrary number of $U(1)$ gauge fields.

Taking the $U(1)$ symmetry into account and choosing the corresponding Killing to be along $\partial_\psi$, i.e. choosing  unknown functions in metric ansatz \eqref{5dU1metric} $R$ and $h$, to be only functions  $\theta$, the equations of motion reduce to
\bea\label{EM5dE}
&& 3R''+6R^{'2}+3R'h'-3p^2e^{-4R}+2\Lambda e^{2R}+6=0\; ,\\
&& 3h''+3h^{'2}+6h'R'+6p^2e^{-4R}+2\Lambda e^{2R}=0\; ,\\
&& 6R^{'2}+6h'R'-3p^2e^{-4R}+2\Lambda e^{2R}{+6}=0\; .
\eea
Subtraction of the first and the third equations in above gives
\be
R''-h'R'=0\quad \Rightarrow\quad R'=\lambda e^h \;,
\ee
where $\lambda$ is an arbitrary integration constant. As in the Einstein-$\Lambda$ case of previous subsection, one can distinguish two cases: $R'\neq0$ or $R'=0$.

\paragraph{$\mathbf{R'\neq 0}$ case.} Eliminating  $h$ in (\ref{EM5dE}) we get
\be
R''+R^{'2}-\frac{\;p^2}{2}e^{-4R}+\frac{\Lambda}{3} e^{2R}+1=0\,.
\ee
One may integrate the above as
\be\label{y-eq}
y'^2+\frac{2\Lambda}{3}y^3+4y^2-Cy+{2p^2}=0\,,
\ee
where $y=e^{2R}$ and $C$ is an integration constant. The solution for $y$ may be written in terms of Weierstrass elliptic function. Alternatively, one can solve the equations more explicitly upon a redefinition in $\theta$ coordinate to obtain the metric
\be\label{EM-Lambda-sol'n}
ds^2=H_{\theta}\bigg[ \mathcal{R}^2  ds_3^2 +\frac{ a^2}{\Delta_{\theta}} (d\theta ^2+  \frac{H_0^3 \Delta_{\theta}^2}{ H_{\theta}^3\Delta_0^2}  \sin^2\theta\cos^2\theta d\psi^2)\bigg],\quad F_{\theta\psi}=\frac{pa^2H_0^{\frac{3}{2}}}{{\mathcal R}^3\Delta_0}\frac{\sin\theta\cos\theta}{H_{\theta}^2}
\ee
where
\bea
&& \Delta_{\theta}=1+\frac{\Lambda a^2}{{6}}\cos^2{\theta}\,,\qquad \;\;\; H_{\theta}=s^2+\cos^2\theta,\\
&&  \mathcal{R}^2=\frac{{a}^2}{1-\frac{\Lambda a^2}{6}(1+3s^2)}\,, \qquad p^2=\frac{2a^4s^2 (1+s^2)(1-\frac{\Lambda}{6} a^2s^2)}{[1-\frac{\Lambda}{6}a^2(1+3s^2)]^3}\,,
\eea
Positivity of $p^2, \mathcal{R}^2$ restricts the range of allowed $a$ for a given $s,\Lambda$: $\Lambda a^2\leq \frac{6}{1+3s^2}$.

For the $\Lambda=0$ case, solution of \eqref{y-eq} takes a simple form,
\be
e^{2R}=\sqrt{\frac{\ell^4}{4}+\frac{p^2}{2}}+\frac{\ell^2}{2} \cos2\theta\;.
\ee
This solution reduces to (\ref{Pure-5d-Einstein-soln}) for $p=0$. Magnetic charge corresponding to this solution is given by
\bea
M=\int F =\sqrt{6}\pi \ell^3 p^{-1}
\eea
One can show that the above may be obtained as the near horizon limit of an  magnetically charged extremal black string.

\paragraph{$\mathbf{R'= 0}$ case.} In this case the solution is of the generic form of AdS$_3\times$X$_2$. If we denote the radius of the AdS$_3$ part by $\ell^2\equiv e^{2R}$ then,
\bea\label{eq3.30}
p^2-{\ell^4}(\ell^2\Lambda+3 )=0 \label{ell-p},\nnr
(e^h)''+(2\Lambda\ell^2+4)e^h=0\label{h-p}\;.
\eea
Eq.\eqref{ell-p} has two (positive) solutions for $\ell^2$ when $\Lambda<0$ and one acceptable solution for $\ell^2$ if $\Lambda>0$.
From \eqref{h-p} we learn that the $X_2$ part of the geometry is a 2d constant curvature space;  the metric for $X_2$ is
\be\label{AdS3X2}
ds_{X_2}^2=\left\{\begin{array}{ccc} R^2(d\theta^2+\sinh^2\theta\psi^2),&\qquad  R^2=\frac{\ell^2}{2\Lambda\ell^2+4}, &\quad \Lambda<-\frac{2}{\ell^2},\ |p|<\ell^2\\ \,\, \\ \ell^2(d\theta^2+\theta^2 d\chi^2), & &  \Lambda=-\frac{2}{\ell^2},\ |p|=\ell^2 \\ \,\,\\ R^2(d\theta^2+\sin^2\theta\psi^2),&\qquad  R^2=-\frac{\ell^2}{2\Lambda\ell^2+4}, &\quad \Lambda>-\frac{2}{\ell^2},\ |p|>\ell^2\end{array}\right.
\ee
As expected, for $p=0$ (and $\Lambda<0$) the solution reduces to AdS$_3\times$H$^2$ discussed in previous subsection. {These three cases are near horizon of EVH black string with flat, spherical and hyperbolic horizon on dS$_5$ or AdS$_5$ background.}

\subsection{Einstein-Dilaton theory}

As the next case we consider Einstein-dilaton theory. The equations of motion for this case are
\be\label{Einstein-Dilaton-5d-eom}
R_{\mu\nu}=2\partial_\mu\Phi \partial_\nu\Phi\,,\qquad \Box \Phi=0\,.
\ee
To solve the above equations we start with the metric ansatz
$$
ds^2=e^{2R} ds_3^2+e^{2f}d\theta^2+e^{2h}d\psi^2\,,
$$
where $R,f$ and $h$ are functions of $\theta,\psi$. In this case, it turns out more convenient to choose the basis such that the coordinate $\theta$ is equal to the dilaton field $\Phi$, i.e. the \emph{linear dilaton} frame (gauge). In the linear dilaton frame $\Phi=\theta$, one may readily solve the equation of motion for the dilaton, and find
$$
f=3R+h\,,
$$
leaving us with two unknown functions $R, h$. Moreover, in this frame, $\partial_\mu\Phi\partial_\nu\Phi=\delta_{\mu,\theta}\delta_{\nu,\theta}$. Among the Einstein equation there are four independent second order partial differential equations of two variables $\theta$ and $\psi$, for these two variables.

Based on these four equations, we could not prove that $SO(2,2)$ invariance implies the extra $U(1)$ isometry. However, we could not find other solutions without the $U(1)$ isometry either. To explore this issue, we checked if the condition for (conformal) Killing equation to have solutions for the background ansatz metric is compatible with the background field equations. Explicitly, if we assume that metric
\be\label{metric-dilaton-theta-ansatz}
ds^2=e^{2R} ds_3^2+e^{6R+2h} d\theta^2+e^{2h}d\psi^2\,,
\ee
has a Killing vector with components in the $\theta\psi$ plane, $\xi=\xi^\theta\partial_\theta+\xi^\psi\partial_\psi$, the condition for (conformal) Killing equation to have solutions reduces to
\bea
 \partial_{\theta} R \partial_{\psi}R \bigg[ \left(\partial^2_{\psi}R-2(\partial_{\psi}R)^2-2\partial_{\psi}R\partial_{\psi}h \right)-e^{6R}\left(\partial^2_{\theta}R-2(\partial_{\theta}R)^2-2\partial_{\theta}R\partial_{\theta}h \right)\bigg]=0\,,
\eea
while from $\theta\theta$ and $\psi\psi$ components of Einstein equations we get
\be
 \left(\partial^2_{\psi}R-2(\partial_{\psi}R)^2-2\partial_{\psi}R\partial_{\psi}h \right)-e^{6R}\left(\partial^2_{\theta}R-2(\partial_{\theta}R)^2-2\partial_{\theta}R\partial_{\theta}h \right)=\frac{2}{3} e^{6R}\,.
\ee
The above two equations are clearly not in agreement, unless $\partial_\theta R$ or $\partial_\psi R$ vanish. The $\theta\psi$ component of Einstein equations, then implies that if $\partial_\theta R$ (or $\partial_\psi R$) vanishes then $\partial_\theta h$ (or $\partial_\psi h$) would vanish too.
That is, a solution of the form  \eqref{metric-dilaton-theta-ansatz} has a Killing direction in $\theta\psi$ plane, only if the metric coefficients do not
have $\theta$ or $\psi$ dependence. Moreover, if one assumes that $R$ is a function of $h$ and its derivatives, i.e.
$R=R(h,\partial_\theta h,\partial_\psi h)$, one may then observe that equations of motion imply that either of $\partial_\psi$ or $\partial_\theta$ are Killing vectors. %
To summarize, this analysis does not rule out solutions without the $U(1)$ isometry, while confirms existence of solutions with the $U(1)$. Moreover, it is straightforward to show that the above analysis regarding the existence of Killing vector extends to Einstein-Maxwell-dilaton theory, with or without $\Lambda$.

Based on the above analysis and the form of Einstein equations, and that we could not find any solution without $U(1)$ isometry, although we could not prove the existence of $U(1)$ for all $SO(2,2)$ invariant solutions to Einstein-dilaton theory, we\emph{ conjecture existence of this extra $U(1)$}. In a similar spirit, we extend this conjecture to the more general case of EMD-$\Lambda$ theory.

Given the above conjecture, assuming existence of a Killing along $\partial_\psi$ direction, here we set about finding solutions to Einstein-dilaton theory.
In this case all metric components are only functions of $\theta$ coordinate, \eqref{Einstein-Dilaton-5d-eom} can be simply integrated out and we get
\be
ds^2=\frac{k e^{\theta}}{2\beta\cosh k\theta}(ds_3^2+\frac{k^2}{4\cosh^2k\theta}\; d\theta^2) +\beta^2 e^{-2\theta} d\psi^2,\quad \Phi'=\frac{\sqrt{3}}{2}\sqrt{k^2-1}
\ee
in the linear dilaton frame.
The above solution may also be written as
\be
ds^2=\frac{R^2\sin2\theta}{\tan^{\alpha}\theta}\left(ds_3^2+d\theta^2  \right)+\beta^2 \tan^{2\alpha}\theta d\psi^2,  \quad  {\Phi}= \frac{\sqrt{3}}{2}\sqrt{1-\alpha^2}\ln(\tan\theta),
\ee
The above solution reduces to the pure Einstein gravity solution for $\alpha=1$ case and has a curvature singularity at $\theta=0$.

We end this part with the other class of solutions of the form AdS$_4 \times_w$I$_1$ (where $I_1$ is an interval):
\be
ds^2=e^{2R} ds_{AdS4}^2+d\theta^2\;,
\ee
where $R$ and scalar field $\Phi$ are functions of $\theta$ and given by real solutions of following equation
\be
(e^{4R})^{''}+12e^{2R}=0,\quad\quad \Phi'=A e^{-4R}\,,
\ee
where $A$ is an integration constant. In presence of a negative cosmological constant $\Lambda$, as in the 4d case discussed in the previous section, we can have AdS$_4 \times_w$I$_1$  solution with linear dilaton background.
\subsection{Einstein-Maxwell-dilaton case}
Again here we adopt the linear-dilaton gauge $\Phi=\gamma \theta$ in which the metric ansatz takes the form \eqref{metric-dilaton-theta-ansatz}. In this gauge the equations of motion for the gauge field considerably simplifies, yielding to\footnote{We are taking the Maxwell-dilaton coupling term $f_{IJ}$ to be one.}
\be\label{gauge-field-EMD-Lambda}
F_{\theta\psi}=p e^{2h}\,,
\ee
where $p$ is  a constant.
One can repeat the analysis for existence of (conformal) Killing vector in $\theta\psi$ plane outlined in the previous subsection and observe that these are not compatible with the equations of motion unless $\partial_\theta R$ or $\partial_\psi R$ are zero. As in the previous subsection, we conjecture  that all $SO(2,2)$ invariant solutions to the EMD-$\Lambda$ theory exhibit an extra $U(1)$.  We construct all such solutions in this section.

In the linear-dilaton gauge with the assumption of extra $U(1)$, the equations of motion for the metric unknowns $R, h$ takes the form
\begin{subequations}\label{EMD-L-5d}\begin{align}
&h''+2 p^2 e^{2h}+\frac{2}{3}\Lambda e^{6R+2 h}=0,\\
&3R''-3 p^2 e^{2h}+2\Lambda e^{6R+2 h}+6 e^{4 R+2 h}=0,\\
&R^{'2}+R'h'+e^{4R+2h}+\frac{1}{3}\Lambda e^{6R+2h}-\frac{1}{2}p^2e^{2h}-\frac{\gamma^2}{3}=0\,.
\end{align}\end{subequations}
Recall that $\Phi=\gamma\theta$ and gauge field is given in \eqref{gauge-field-EMD-Lambda}.
One can easily check that (\ref{EMD-L-5d}.c) is the integrability condition for the other two equations, i.e. the above three equations for two unknowns $h, R$ are compatible with each other.
As a check, we observe that for the special case of $k=0$ the above equations reduce to those of Einstein-Maxwell-$\Lambda$ given in \eqref{EM5dE}. (For the latter please note that \eqref{EM5dE} is written for a different ansatz metric.) It may also be checked that for $p=0,\ \Lambda=0$ we recover the  Einstein-dilaton case discussed in previous subsection. Hereafter we will hence assume $\gamma, p$ are both non-zero.

We first note that, the above equations do not have a constant $R$ solution (except for $\gamma=0$ case). That is, for generic EMD-$\Lambda$ theory we do not have AdS$_3\times X_2$ solution. One may also check that we do not have AdS$_4\times_w$I$_1$ solution (except for $p=0$ case). These results are true for any sign and value of $\Lambda$.

For the EMD case with $\Lambda=0$, one can solve the above equations to obtain $R, h$:
\be
e^{2R}=\frac{p k_2 \cdot\cosh(k_1\theta+\theta_0)}{\sqrt{2}\cosh k_1k_2\theta},\quad   e^{2h}=\frac{k_1^2}{2p^2\cosh^2(k_1 \theta+\theta_0)},\quad k_2=\sqrt{1+\frac{4}{3}\gamma^2k_1^2\;}\,,
\ee
(note that the dilaton field is $\Phi=\gamma \theta$.)
We note that the solutions for the special case of $p=0$ (the Einstein-dilaton case of previous section) cannot be obtained as a simple $p\to 0$ limit of the above solution.\footnote{In order to recover this solution we need to take a singular limit of the above solution, while keeping the integration constant associated with the shift in $\theta$, $\theta_0$ and take an appropriate simultaneous $\theta_0\to\infty, p\to 0$ limit. }

\paragraph{EMD theory with Maxwell coupling $f=e^{k\Phi}$.} Equations \eqref{EMD-L-5d} was written for a theory with cosmological constant but without coupling of dilaton to Maxwell field. i.e. $f=1$ in the notation of \eqref{5d-EOM}. In usual supergravity theories, however, there is a coupling of the form $f=e^{k \Phi}$ for specific values of $k$. In this case the equations of motion are of the form:
\begin{subequations}\label{EDM-f-EOM}\begin{align}
R_{\mu\nu}=2\partial_{\mu}\Phi\partial_{\nu}\Phi+e^{k\Phi} (&2F_{\mu\lambda} F_{\nu}^{\ \ \lambda}-\frac{1}{3}g_{\mu\nu} F_{\alpha\beta} F^{\alpha\beta}), \\
 \frac{1}{\sqrt{-g}} \partial_{\mu}\left(\sqrt{-g} \partial^{\mu} \Phi\right) &=\frac{k}{4}e^{k\Phi}F_{\mu\nu}F^{\mu\nu}, \\
\partial_{\mu}\bigg[ \sqrt{-g} e^{k\Phi} F^{\mu\nu}\bigg]&=0.
\end{align}\end{subequations}
It turns out convenient to adopt the following parametrization for the metric ansatz
\be
ds^2=e^{2R} ds_3^2+ e^{6R+2h}d\theta^2+e^{2h} d\psi^2\,,
\ee
where $R, h$ are functions of $\theta$. The solution to gauge field equations of motion is easily obtained as
\be
F_{\theta\psi}=p e^{-k\Phi+2h}\,,
\ee
where $p$ is an integration constant. The other equations of motion are then obtained as
\begin{subequations}\begin{align}
&R''+2e^{4R+2h}-\frac{2p^2}3 e^{-k\Phi+2h}=0\,,\\
&h''+\frac{4p^2}3 e^{-k\Phi+2h}=0,\\
&\Phi''-\frac{p^2k}{2}\ e^{-k\Phi+2h}=0\,,\\
&R^{'2}+R'h'+e^{4R+2h}-\frac{p^2}{3}e^{-k\Phi+2h}-\frac{\Phi'^2}{3}=0\,.
\end{align}\end{subequations}
(The last equation is not independent of the first three.) One may solve the above equations to obtain
\be\begin{split}
\Phi=-\frac{3k}{8}h+b\theta\,,\qquad e^{2R+h}=\frac{a}{2\cosh a(\theta+\theta_0)}\,,\qquad
e^{h-\frac{k}{2}\Phi}=\frac{c}{m\cosh c\theta}\,,
\end{split}\ee
where
$$
m^2=p^2(\frac43+\frac{k^2}{4})\,,\qquad a^2(1+\frac{3k^2}{16})=c^2+\frac43b^2\,,
$$
and $b, c$ and $\theta_0$ are integration constants.

\subsection{$U(1)^3$ gauged supergravity}\label{section-u1-3}

Although the $U(1)^3$ gauged SUGRA is not of the form of EMD theory, for completeness we also present solutions of this theory with local $SO(2,2)$ symmetry. The bosonic part of this SUGRA has three $U(1)$ gauge fields and two scalars with the action
\bea\label{U(1)3-SUGRA}
&& S=\frac{1}{16\pi G}\int d^5x\bigg[R-\frac12\sum_{i=1}^3\ {X_i}^{-2} F_i^2 -2(\partial\Phi_1)^2-2(\partial\Phi_2)^2 \nonumber\\
&&\hspace{32mm} +\frac{4}{\ell^2}\sum_{i=1}^3 X_i^{-1}+2F_1\wedge F_2\wedge A_3\bigg]
\eea
where
\be\label{Xi}
X_1=e^{-\frac{2}{\sqrt 6}\Phi_1-\sqrt{2}\Phi_2}\,,\qquad X_2=e^{-\frac{2}{\sqrt 6}\Phi_1+\sqrt{2}\Phi_2}\,,\qquad
X_3=e^{\frac{4}{\sqrt 6}\Phi_1}\,.
\ee
This theory may be obtained from a reduction of 10d IIb SUGRA over an $S^5$ \cite{ten-people}.
Therefore, all solutions to this 5d SUGRA have a 10d uplift.

The vacuum solution to this theory (when the gauge fields  and scalars turned off) is an AdS$_5$ geometry of radius $\ell$. The most general black hole solution to this theory has been constructed in \cite{Wu-black-hole} and has six parameters, a mass parameters, two angular momenta and three electric charges.

For the general metric ansatz of the form
\bea\label{U(1)3-metric-ansatz}
ds^2=e^{2R(\theta)}ds_3^2+e^{2f(\theta)}d\theta^2+e^{2h(\theta)}d\psi^2\,,
\eea
the $SO(2,2)$ invariant solution to gauge field equations of motion takes the form
\be\label{U(1)3-gauge-field}
F^{(i)}_{\theta \psi}=p_i X_i^{2}e^{f(\theta)+h(\theta)-3R(\theta)},
\ee
where $p_i$ are constants to be fixed by the equations of motion for metric and scalars. Note that since all three gauge fields have only $F_{\theta\psi}$ components, the Chern-Simons term on the action does not contribute to the equations of motion for the class of solutions of our interest.

Here we discuss three class of solutions to the $U(1)^3$ gauged SUGRA theory:

\textbf{I.} A general four parameter class of solutions to this theory with $SO(2,2)\times U(1)$ is obtained to be
\be\label{U(1)3-metric}
ds^2=H_{\theta}\Bigg[ \mathcal{R}^2   ds_3^2 +\frac{ a^2 }{\Delta_{\theta}} (d\theta ^2+  \frac{H_0^3}{H_{\theta}^3}\frac{\Delta_{\theta}^2}{\Delta_0^2}\; \sin^2\theta\cos^2\theta d\psi^2)\bigg],
\ee
where
\bea
\Delta_{\theta}=(1-\frac{a^2}{\ell^2}\cos^2{\theta})\,,\quad H_{i}={\cos^2{\theta}+s_i^2},\quad H_{\theta}=H_1^{\frac{1}{3}} H_2^{\frac{1}{3}} H_3^{\frac{1}{3}} ,\quad
\eea
and scalar fields are given in terms of $X_i$ which are
\bea
X_i=\frac{H_{\theta}}{H_i}\,.
\eea
The constants $p_i$ and $R^2$ are related to $a$ and $s_i$ as
\bea
&&\mathcal{R}^2= \frac{a^2}{1+\frac{a^2}{\ell^2}(s_1^2+s_2^2+s_3^2+1)},\\
&&p_i^2=\frac{2 a^4 s_i^2(s_i^2+1)(1+ \frac{a^2}{\ell^2}s_i^2)}{[1+\frac{a^2}{\ell^2}(1+s_1^2+s_2^2+s_3^2)]^3},
\eea
We comment that for the ``two charge case'', when e.g. $s_3=0$, the above solution reproduces the one discussed in \cite{EVH-3}. (To compare the two one should consider the 10d uplift of the above solutions and make appropriate $\theta$-coordinate transformation.)

\textbf{II.} There is another class of solutions with $R(\theta)=const$. In this case equations of motion imply that the scalars are also constant. Therefore, for this class of solutions the rest of equations of motion reduces to that of an Einstein-Maxwell-$\Lambda$ theory with three $U(1)$ gauge fields. That is, we have solutions of the form AdS$_3\times X_2$, where $X_2$ is a maximally symmetric space, the same as what we had in \eqref{eq3.30} and \eqref{AdS3X2}.

\textbf{III.} The $U(1)^3$ theory of course admits an AdS$_5$ solution, which comes with $F^{(i)}=0$ and $X_i=1$. The AdS$_5$ metric in AdS$_3$ slicing is of the form
\be\label{AdS5-AdS3-slicing}
ds^2=\ell^2\cosh^2\theta ds_3^2+\ell^2d\theta^2+\ell^2\sinh^2\theta d\psi^2\,.
\ee

\paragraph{Consistent truncations.}  There exit various consistent truncations of the $U(1)^3$ gauged SUGRA theory \eqref{U(1)3-SUGRA} to the theories discussed in previous sections. Given these truncations one can check and reproduce
some of the solutions to these theories which were all classified in the previous sections.\footnote{Note that although the solutions of the $U(1)^3$ theory given here cover a large class of its solutions with $SO(2,2)\times U(1)$ symmetry, our solutions may not classify all such solutions of the $U(1)^3$ theory.} Here we discuss these truncations.
\begin{itemize}
\item \textbf{Truncation to Einstein-$\Lambda$ theory.} It is readily seen that  $X_i=1, F_{\mu\nu}^i=0$ gives a consistent truncation of $U(1)^3$ theory to Einstein-$\Lambda$ theory with $\Lambda=-\frac{6}{\ell^2}$. Moreover, one can observe that the class of solutions given in \eqref{U(1)3-metric} for $s_i=0$ reduces to \eqref{5d-Lambda-metric}.
In addition, one can also see that the AdS$_3\times X_2$ solutions of the $U(1)^3$ theory reduces to similar solutions in Einstein-$\Lambda$ theory. Therefore, one can recover all solutions discussed in Einstein-$\Lambda$ theory from the truncation of solutions of the $U(1)^3$ theory we discussed here.

One may also take the $\ell\to \infty$ limit and further reduce the above Einstein-$\Lambda$ theory to pure Einstein gravity and observe that all solutions of the section 5.1 can be reproduced from the $U(1)^3$ theory solutions we presented here.

\item \textbf{Truncation to Einstein-Maxwell-$\Lambda$ theory.} The $A^{(i)}_\mu=A_\mu$ which equates the three $U(1)$ gauge fields constitutes a consistent truncation if $X_i$ are also equal to one. This will therefore consistently reduces the $U(1)^3$ theory to an Einstein-Maxwell-$\Lambda$ theory with $\Lambda=-\frac{6}{\ell^2}$. One may apply the same truncation to the solutions above by setting $s_i=s$. In this case $H_i=H$ and $X_i=1$, and the solution \eqref{U(1)3-metric}  reduces to \eqref{EM-Lambda-sol'n}. In a similar manner, one can reproduce AdS$_3\times X_2$ solutions of the Einstein-Maxwell-$\Lambda$ theory from the solutions of $U(1)^3$ theory we discussed above. In other words, all of solutions discussed in section 3.3 can be reproduced from $U(1)^3$ solutions discussed here.

One may also take the $\ell\to \infty$ limit and further reduce the above Einstein-Maxwell-$\Lambda$ theory to Einstein-Maxwell gravity.

\item \textbf{Truncation to Einstein-Maxwell-Scalar/Dilaton theory.} One can show that $\phi_2=0$ or $X_1=X_2$ and $A_1=A_2=0$ provides a consistent truncation of the $U(1)^3$ theory to Einstein-Maxwell-Scalar theory with the action
\be\label{U(1)3-->EMS}
{\cal L}= R- \frac14 e^{-{8}\phi/\sqrt6} F_{\mu\nu}^2-2(\partial\phi)^2+\frac{4}{\ell^2} (2 e^{{2}\phi/\sqrt6}+e^{-{4}\phi/\sqrt6})\,.
\ee
It is also possible to further restrict the theory by sending $\ell\to\infty$, where the theory takes the form of EMD of the form discussed in section 3.5 with gauge field-dilaton coupling with $k=-8/\sqrt6$. One may readily observe that the solutions of the $U(1)^3$ we have given here for $s_1=s_2=0$ reproduce a class of solutions of the EMD of previous section.

\end{itemize}

\section{General $d$-dimensional vacuum solutions}\label{sec-4}

In the previous section we gave a (more or less) complete classification of four and five dimensional $SO(2,2)$ invariant solutions (see the discussion section for more detailed summary). In this part we give three classes of $SO(2,2)$ invariant solutions to higher dimensional pure Einstein gravity. In $d\geq 6$ dimensions the equations of motion in general reduce to non-linear \emph{partial} differential equations on the $d-3$ dimensional space. There is no standard procedure to classify all solutions to these equations. Here we construct three class of such solutions which are those with $SO(2,2)\times U(1)^{d-4}$, $SO(2,2)\times SO(d-3)$ and a family of $SO(2,2)\times U(1)^{n}$ ($n=\frac{d-3}{2}$ in odd dimension) isometry.
\begin{itemize}
\item  Imposing the extra $d-4$ $U(1)$ symmetries reduces the partial differential equations to ordinary second order differential equations, whose solutions can be uniquely specified if we assume that the $d-4$ dimensional part of metric is smooth. Explicitly, let us consider the ansatz
\be
ds^2=e^{2R}\left(-r^2dt^2+\frac{dr^2}{r^2}+r^2d\phi^2 + d\theta^2\right) +\sum_{i=1}^{d-4}e^{2h_i}d\psi_i^2+\sum_{i=1}^{d-4} f_i\ d\theta d\psi_i
\ee
where $R,\ h_i$ and $f_i$ are real functions of $\theta$. One can always remove off-diagonal $f_i$ terms by a shift in $\psi_i$. We will henceforth work in such a frame.  For this class of solutions the $d-3$ dimensional part of the geometry is a compact, finite volume and smooth surface with topology of a solid torus; i.e. at generic constant $\theta$ surface we find a $d-4$ dimensional torus $T^{d-4}$.

It is easy to check that vacuum Einstein equations does not admit any solution for $d\leq 4$. For $d>4$ however we get following solutions
\bea
\label{sol}
&&e^{2R}=A^2 \sin(\theta) \cos(\theta) \left(\tan\theta\right)^{a},\;\;\; e^{2h_i}=B_i^2 (\tan\theta)^{c_i} ,
\eea
where
\be
 2a+\sum_{i=1}^{d-4} c_i=0,\;\; 2a^2+\sum_{i=1}^{d-4} c_i^2=6\,.
\ee
One can readily observe that for $d=4$ the above have not real solutions for $a,c$, compatible with the results of section 2, and for  $d=5$ it reproduces the solution in \eqref{Pure-5d-Einstein-soln}.

One can check that the Kretschmann invariant  blows up at  $\theta=\pi/2$ and hence we have a curvature singularity. Nonetheless, one can show that upon the reduction of $d$ dimensional Einstein gravity to three dimensions we obtain an AdS$_3$ gravity with a finite Newton constant and finite AdS$_3$ radius.


\item As the next geometry with the maximal rotation symmetries, one may easily show that there are no solutions with $SO(2,2)\times SO(d-2)$ isometry (i.e. an AdS$_3\times$S$^{d-3}$. The next case with smaller rotation group is  the solution with $SO(2,2)\times SO(d-3)$ isometry. To this end, we start with the metric ansatz
\be
\label{Sd4}
ds^2=e^{2R(\theta)} (ds_3^2 +d\theta^2)+ e^{2f(\theta)} d\Omega_{d-4}^2.
\ee
Equations of motion for above metric ansatz are given by
\bea\label{geneq}
&&R''+2R^{'2}+(d-4)R'f'+2=0,\\
&&(d-4)(f''+f'^2-4R'f')-{6R^{'2}}-{6}=0,\\ \label{geneq3}
&&(d-4)(d-5)(f^{'2}-e^{2R-2f})+6(d-4)R'f'+6R^{'2}+6=0.
\eea
Although we could not find an explicit solution for the above equations, one may show that the above equations do have a solution  for which $e^R$ is an even function of $\theta$ and $e^f$ is an odd function. The series expansion for the solutions is obtained to be
\bea
&& e^R=\ell\bigg(1-\frac{\theta^2}{d-3}+\frac{\theta^4}{6(d-3)^2}-\frac{(d^2-27d+116)\theta^6}{90(d-4)(d-3)^3(d+1)}\cr
&&\quad+\frac{(d^4+572d^3-7789d^2+33448d-44592)\theta^8}{2520(d-4)^2(d-3)^4(d+1)(d+3)}+\cdots \bigg),\cr\cr
&&e^f=\ell\bigg(\theta-\frac{(d-7)\theta^3}{3(d-4)(d-3)}+\frac{(d-1)^2 \theta^5}{30 (d-4)^2(d-3)^2}\cr
&&-\frac{(d^4-128d^3+1206d^2-4912d+8153)\theta^7}{630(d-4)^3(d-3)^3(d+1)}+\cdots\bigg)
\eea
where for $d=5$ we recognise Taylor expansion of $\cos\theta$ and $\tan\theta$.\footnote{We note that as in the $d=5$ case and as discussed in \cite{Flat-reduction} the solutions in the above class are not smooth and have curvature singularity.}

\item In odd dimensions $d=2n+3$, one can find another class of vacuum solutions with $SO(2,2)\times U(1)^n$ isometry.  This class of solutions which are specified by $n$ free parameters $a_i$ are given by
\be\label{NH-general-Kerr-EVH}%
ds^2 = \cos^2\theta ({\cal R}^2 ds_3^2+d\theta^2)
 + \sin^2\theta \sum_{i=1}^{n}
\left[ a_i^2 d\mu_i^2 + a_i^2 \mu_i^2\left(1+\tan^2\theta {\mu_i^2}
\right)d\psi_i^2\right] ,%
\ee%
\be
{\cal R}^2= \left(\sum_{i=1}^{n} \frac{1}{a_i^2}\right)^{-1}\,,\qquad \sum_{i=1}^n \mu_i^2=1\,.
\ee
It is readily seen that the above solution in $d=5$ or $n=1$ reduces to \eqref{Pure-5d-Einstein-soln}. This class of solutions are related to the near horizon limit of odd dimensional EVH Myers-Perry black holes \cite{EVH-2}. (In the class of MP black holes in odd dimensions, EVH black holes are those when only one of the angular momenta vanishes.)

\end{itemize}

\section{Conclusion}\label{sec-5}

In this work we set about classifying all $SO(2,2)$ invariant EMD-$\Lambda$ and the STU gauged supergravity theories in four and five dimensions. To the EMD-$\Lambda$ case we gave a complete classification. Note that while in 4d and 5d EM-$\Lambda$ our classification was complete, in the 5d EMD-$\Lambda$ case, our classification is complete only if we assume the extra $U(1)$ isometry. For the STU gauged supergravities we provided a large class of such solutions, but could not show that these solution are exhaustive. In the 5d $U(1)^3$ gauged supergravity case we discussed various truncations of the theory to EMD-$\Lambda$ theories and observed that the class of solutions we have given already exhausts all the solutions of the EMD-$\Lambda$ theory in the truncating limit. This could be viewed as evidence supporting the view that, despite the lack of a proof, our class of solutions to $U(1)^3$ theory already exhausts all the solutions in the class of our interest.

\begin{center}
    \begin{tabular}{ | p{6mm} | p{21mm}| p{21mm} | p{21mm} | p{21mm} | p{21mm} | p{30mm} |}
    \hline
     & Vacuum Einstein  & E-$\Lambda$ & E-M-$\Lambda$ & E-D & E-M-D  & Gauged SUGRA\\ \hline
    d=4 & No solution exists & AdS$_4$ is the only
 solution & No gauge field in 4d is allowed; AdS$_4$ is the only
 solution   &   Complete classification by \eqref{ads3s1}, \eqref{NH4dEVH}& Complete classification by \eqref{ads3s1}, \eqref{NH4dEVH} &A set of solutions to $U(1)^4$ given in \eqref{2.16}, \eqref{2.17}.\\ \hline
    d=5 & U(1)-theorem; (3.8) exhausts all solutions. & U(1)-theorem; (3.13), (3.15), (3.16) classify all solutions & U(1)-theorem; (3.25), (3.28) give complete Classification & (3.36), (3.37) give complete classification, \emph{assuming the U(1)} &(3.42), (3.47) are exhaustive \emph{assuming the U(1)} &(3.52) gives a large class of solutions to U(1)$^3$\\ \hline
    d$>$5 &(4.2), (4.8), (4.9), specific solutions &  &   & &  &\\
    \hline
    \end{tabular}
\end{center}

One of the motivations for this study was related to EVH black holes/rings and the conjecture that any EVH black hole/ring in the near horizon limit produces a (locally) AdS$_3$ throat, while the converse is not necessarily true: that any solution with $SO(2,2)$ isometry is not necessarily coming from the near horizon limit of an EVH black hole/ring. This conjecture was proved in 4d case in \cite{EVH/CFT}. Extending the EVH uniqueness theorem to 5d and proving that any 5d EVH black hole/ring solution to EMD theory has indeed an AdS$_3$ throat in its near horizon limit is the problem we would tackle in our upcoming publication. Moreover, as discussed in this work, most of the solutions we gave in four and five dimensions are indeed coming from the near horizon limit of known EVH black holes/rings. In particular, we have checked that the solutions given in section 3.6 for $U(1)^3$ theory may be obtained from the near horizon limit of the most general black hole solutions to this theory constructed by Wu \cite{Wu-black-hole}. The Wu black holes constitute  a six parameter family of solutions, specified by a mass, two spins and three (electric) charges, denoted by $(m;a,b;s_1,s_2,s_3)$ in notation adopted in \cite{Wu-black-hole}. The EVH black holes in this class then come with one of the spin parameters, say $b$ is equal to zero, together with the ``extremality condition'' $2m=a^2$.

We also discussed some classes of higher dimensional vacuum solutions. A class of them with $SO(2,2)\times U(1)^n$ where $n=\frac{d-3}{2}$ for odd $d$, is coming from the near horizon limit of EVH MP black holes \cite{EVH-2}. While we expect the other solutions we discussed here to be also related to higher dimensional vacuum solution EVH black holes, we have not found such explicit black hole solutions in the literature. (As discussed in \cite{EVH-Ring} in the 5d case in the near horizon limit of EVH black rings and MP black holes become the same geometry, while one can still try to trace ring vs. hole question analysing near-EVH geometry.)

In the 5d case, we proved a ``rigidity theorem'' stating that in the Einstein-Maxwell-$\Lambda$ theory $SO(2,2)$ invariance implies an extra $U(1)$ isometry. Although we could not prove (due to technical difficulty of dealing with highly complicated partial differential equations) the same statement for the EMD-$\Lambda$ theory, we expect this to be true for this general case, as well as the $U(1)^3$ gauged supergravity.

All our $SO(2,2)$ invariant solutions come with an AdS$_3$ factor and a ``transverse'' $d-3$ part which has a finite volume (or can be made compact). One may then reduce the $d$ dimensional gravity theory to obtain a 3d Einstein AdS$_3$ gravity. This reduction has been carried out in the appendix. In the solutions we discussed here the AdS$_3$ part generically comes with a warp factor which has zeros and one can show that the whole geometry has curvature singularity at points where the warp factor vanishes. De spite having curvature singularity, the modes in the AdS$_3$ gravity obtained upon the reduction are smooth. This implies that these are ``good'' singularities in the sense used in \cite{Gubser:2000nd}. Given the curvature singularity one may examine whether the higher derivative corrections to gravity action can remove the singularity. Such an analysis has been carried out in \cite{Hossein-HD}.

Finally, given the AdS$_3$ factor, one may put forward the EVH/CFT proposal, stating that low energy excitations on the geometries we discussed here is dual to a 2d CFT associated with the AdS$_3$ factor. Studying various aspects of this proposal is an interesting question we postpone to future works.

\appendix
\section{Dimensional reductions}
Let us write the metric in form
\be
ds^2=e^{2R}ds_3^2+h_{ab}dx^adx^b,
\ee
where $R$ is a function of internal coordinates $x_a$, $ds_3^2$ is the line element on three dimensional space-time and $h_{ab}$ is the metric on the ''internal" $d-3$ dimensional part. Non-zero components of Ricci tensor are given by
\be
{\mathcal R}_{\mu\nu}={\mathcal R}_{\mu\nu}^{(3)}-g_{\mu\nu}^{(3)}\left( \nabla_i\nabla^i R+3\nabla_iR\nabla^iR\right) e^{2R},\quad {\mathcal R}_{ij}={\mathcal R}_{ij}^{(d-3)}-3\nabla_i\nabla_j R-3\nabla_iR\nabla_jR\;,
\ee
and Ricci scalar turns out to be
\be
{\mathcal R}=e^{-2R}{\mathcal R}_{(3)}+{\mathcal R}_{(d-3)}-6\nabla_i\nabla^i R-12\nabla_i R\nabla^i R
\ee
where we used indexes $i$ and $j$ for the internal space.

\subsection{Reduction over the AdS$_3$ down to $d-3$ Euclidian space}
One may reduce the theory on the AdS$_3$ part to obtain a $d-3$ dimensional Euclidian theory. This $d-3$ theory is in fact the theory whose solutions would classify all $SO(2,2)$ invariant solutions we discussed in the work. Here we discuss reduction of Einstein-Hilbert action, 4d Einstein-dilaton theory and 5d EMD-$\Lambda$ theory separately.

\paragraph{Reduction of Einstein-Hilbert action} down to $d-3$ leads to
\be
S= \int \sqrt{h} e^{3R}\left(  {\mathcal R}_{(d-3)}+6(\nabla R)^2-6e^{-2R}\right)d^{d-3}x
\ee
In particular when the internal space is in form (\ref{Sd4}), above action will reduce to following one dimensional action
\bea
S=\int e^{2R+(d-4)f}\bigg((d-4)(d-5)f^{'2}+6R^{'2}+6(d-4)f'R'\qquad \cr+(d-4)(d-5)e^{2R-2f}-6\bigg)d\theta
\eea
It is easy to check that equations (\ref{geneq})-(\ref{geneq3}) can be obtained from above action.

\paragraph{Reduction of four dimensional EMD-$\Lambda$ theory.}
As discussed gauge fields do not contribute to $SO(2,2)$ invariant geometries. Therefore, one may simply consider Einstein-Dilaton/Scalar theory, turning off the gauge fields.
It is straightforward to check that equations (\ref{4d-scalar-EOM})-(\ref{4d-scalar-EOM2}) can be derived from the following one-dimensional action
\be
S= \int e^f\left(3f^{'2}-4\Phi^{'2}-V_{eff}\right)d\theta\;, \quad V_{eff}=2\beta^2e^fV(\Phi) +12 \beta^2
\ee
\paragraph{Reduction of five dimensional EMD-$\Lambda$ theory.}
Using diffeomorphisms to write two dimensional metric in diagonal form $h_{ij}=e^{2A}\delta_{ij}$,  equations of motion for metric (\ref{genericmetric}), scalars and gauge fields (\ref{gaugefield5dim}) can be derived from the following two dimensional action
\be
S=\int \bigg( {\mathcal G}_{rs}\delta^{ij}\partial_{i} X^{r} \partial_j X^s -V_{eff}(X)\bigg) d\theta d\psi \, , \quad r, s= 1,\cdots, N_{\Phi}+2\;,
\ee
where $N_{\Phi}$ is number of scalar fields. Two dimensional scalar fields $X^r$ and effective potential $V_{eft}$ are defined by
\be
X^1=R\;,\;\;X^2=A\;,\;\; X^{2+k}=\Phi^k\;,\;\;V_{eff}=6e^{2A}-2e^{2A-4R}f_{IJ}p^Ip^J+e^{2R+2A}V(\Phi)\,.
\ee
Metric ${\mathcal G}$ is defined by
\be
{\mathcal G}_{rs} =
 \begin{pmatrix}
   3e^{3X^1} & 3e^{3X^1} & 0& \cdots&0 \\
  3e^{3X^2} & 0 & 0& \cdots&0 \\
  0 & 0 & & &\\
   \vdots& \vdots &  &  e^{3X^1} G_{mn} \\
   0 & 0 &  &  &
 \end{pmatrix}.
\ee
\subsection{Reduction of $d$ dimensional space down to 3d space}
Starting from Einstein-Hilbert action in $d$-dimensions
\be S=\frac{1}{16\pi G_d}\int\sqrt{-g_d} {\mathcal R_{(d)}} d^dx \ee
and integrating over internal space coordinate we get
\be
S=\frac{1}{16\pi G_3}\int \sqrt{-g_{(3)}}\left( {\mathcal R}_3-2\Lambda  \right) d^3x\;.
\ee
For metric ansatz  (\ref{Sd4}), three dimensional Newton constant $G_3$ and cosmological constant $\Lambda$ are given by
\be
G_3^{-1}=\ell^{d-3}S_{d-4}c_{d} G_d^{-1} ,\quad \Lambda={-6}\ell^{-2}\;.
\ee
where $c_{d}$ is a number of order unity and $S_d$ is area of unit d-sphere. For solution (\ref{NH-general-Kerr-EVH}) $G_3$ turns out to be
\be
G_3^{-1}=\ell^{d-3} (2\pi)^{d-4} G_d^{-1}
\ee
 For the the five dimensional example $U(1)^3$ we get
\be
G_3^{-1}=\pi a^2H_0^{\frac{ 3}{2}}\Delta_0^{-1} G_5^{-1},
\ee
and for the four dimensional example $U(1)^4$ it turns out to be
 \be
 G_3^{-1}=\sqrt{2} \mathcal{R} G_4^{-1}\,.
 \ee

\bibliographystyle{plain}

\end{document}